\documentstyle[aps,epsfig]{revtex}
\newcommand{\om}{\Omega_{\rm M}}
\newcommand{\ola}{\Omega_\Lambda}
\begin{document}
\twocolumn

\title{The Dyer-Roeder distance-redshift relation in inhomogeneous universes}
\author{E. M\"ortsell\thanks{Electronic address: edvard@physto.se}}
\address{Department of Physics, Stockholm University, \\
         S--106 91 Stockholm, Sweden}
\maketitle

%{\bf Received...}
%==========
\begin{abstract}
Using Monte-Carlo methods, we determine the best-fit value of the 
homogeneity parameter $\alpha$ in the Dyer-Roeder distance-redshift relation
for a variety of redshifts, inhomogeneity models and cosmological parameter 
values. The relation between $\alpha$ and the 
fraction of compact objects, $f_{\rm p}$, is found to be approximately 
linear. This relation can be parametrized with reasonable accuracy for all
cases treated in this paper  
by $1-\alpha =a\cdot f_{\rm p},$ where $a\approx 0.6$.
\end{abstract}
\vspace{3mm}

%==========
\section{Introduction}\label{sec:intro}
Assuming the validity of the cosmological principle, i.e., that the Universe is isotropic 
and homogeneous, we can apply the Robertson-Walker (RW) metric to the field equations of 
general relativity to derive a
relation of cosmological distance measures to redshift for various cosmological parameter 
values. However, we know that our Universe is very far from homogeneous on scales smaller 
than galaxy clusters. It is generally assumed that this does not affect the large scale
expansion rate of the Universe. Still, inhomogeneities will affect measured distances 
through the effect of gravitational lensing. There will be different amounts of matter
along different lines-of-sight, causing different amounts of focusing of the light-rays.
It is not possible to obtain exact solutions of the field equations for general 
inhomogeneous models, thus one is referred to numerical simulations to
compute gravitational lensing effects on distance measurements. 

Sometimes, we would like to
be able to use simpler methods to compute at least approximate distances.
The Dyer-Roeder (DR) distance-redshift relation \cite{dr} assumes that
the expansion rate of the
Universe is governed by the total matter density whereas the focusing of light is
only affected by a fraction $\alpha$ of the total matter density.
The DR distance 
thus contains an additional parameter, namely the homogeneity-parameter, $\alpha$. 
The approximation should be fair
if a fraction $1-\alpha$ of the matter density is in very compact objects and
the light-ray travels far from all matter accumulations, i.e., if  
one can neglect
the effect of gravitational lensing.

In this paper we investigate properties of the DR distance-redshift relation
by comparing with
numerical simulations of the distance-redshift relation
in inhomogeneous universes, including the effect from gravitational lensing. 
More specifically,
we compute the best fit value of the homegeneity-parameter $\alpha$ for different
cosmologies and inhomogeneity models. These values can be used, e.g., 
with some of the publicly available 
routines for computing cosmological distances \cite{helbig}. 

In an earlier study, Tomita \cite{tomita} has found the best-fit value of $\alpha$ 
to be one in most cases, with a dispersion in $\alpha$ dependent on the cosmological model, 
the redshift and separation of light-rays.

%==========
\section{Method and results}\label{sec:method}
\begin{figure}[t]
  \centerline{\hbox{\epsfig{figure=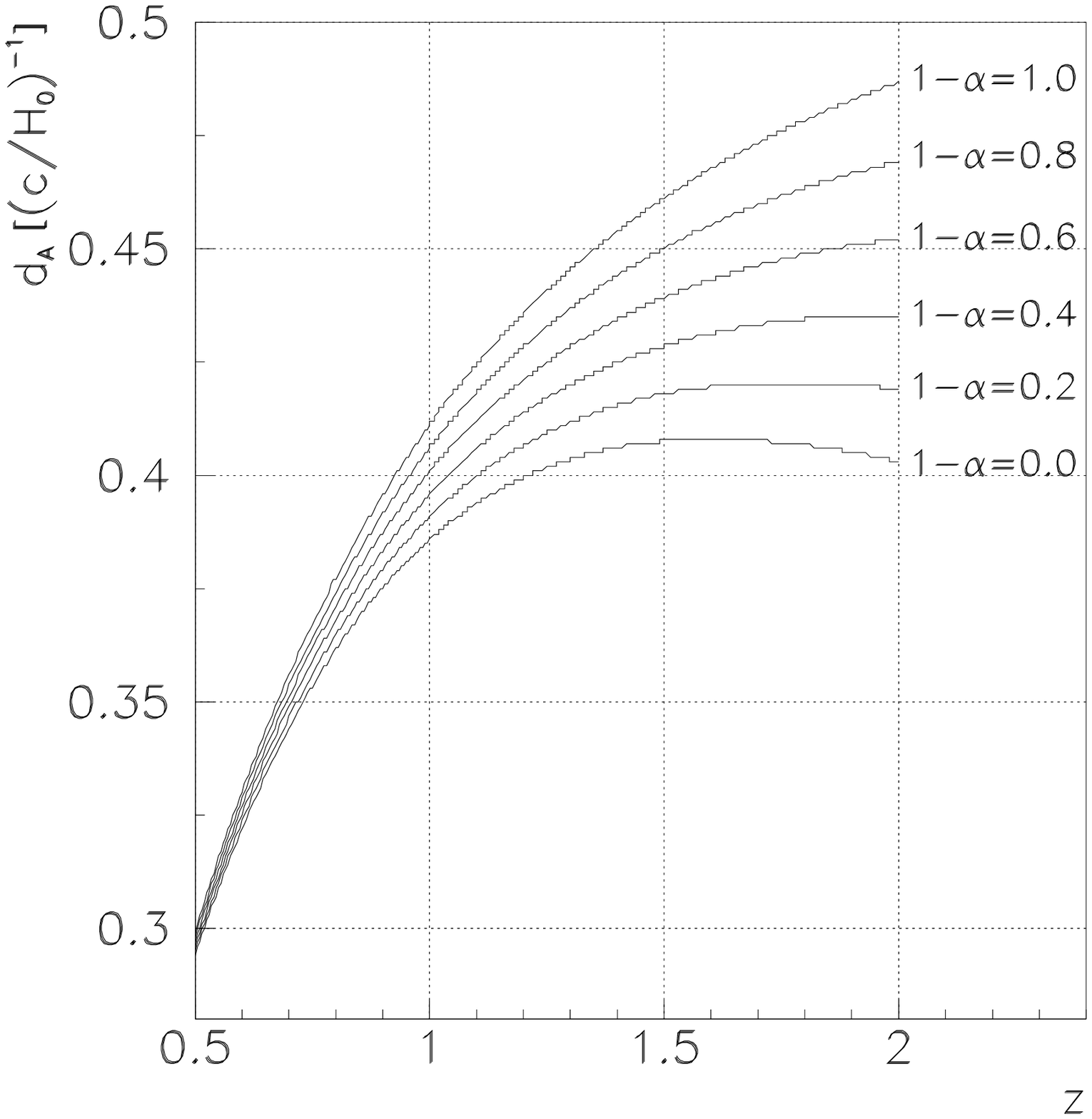,width=0.5\textwidth}}}
  \caption{The Dyer-Roeder angular-diameter distance in units of the Hubble-length for
    different values of the homogeneity-parameter $\alpha$ and 
    $(\om ,\ola)=(0.3,0.7)$.}
  \label{fig:dr} 
\end{figure}

In Fig.\ref{fig:dr}, we have plotted the DR angular-diameter 
distance in units of the Hubble-length for
different values of the homogeneity-parameter $\alpha$. The larger the value of
$\alpha$, the larger the amount of focusing of the light-rays. Thus, the 
angular-diameter distance decreases monotonically with $\alpha$, i.e.,
it increases with the "inhomogeneity-parameter", $1-\alpha$.
Note also that there is a simple relation between the luminosity distance, 
$d_{\rm L}$, 
and the angular-diameter distance $d_{\rm A}$, $$d_{\rm L}=(1+z)^2\cdot d_{\rm A}.$$

\begin{figure}[t]
  \centerline{\hbox{\epsfig{figure=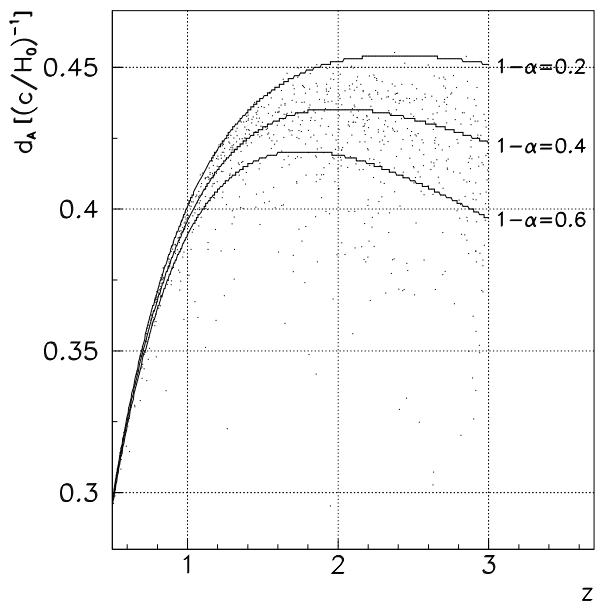,width=0.5\textwidth}}}
  \caption{The dispersion in $d_{\rm A}$ for $(\om ,\ola)=(0.3,0.7)$
    with a fraction 0.6 of the matter density 
    in galactic dark matter halos and a fraction 0.4
    in compact objects.}
  \label{fig:sigma} 
\end{figure}

Using the simulation package SNOC \cite{bergstrom}, we have used light-ray tracing 
to obtain angular-diameter distances with different inhomogeneity models and 
different values of the cosmological parameters. Gravitational lensing effects
are calculated by integrating the geodesic deviation equation through a number of 
consecutive cells between the observer and the source. In each cell, we can specify
the matter distribution governing the deviation. For more details of the method,
originally proposed by Holz and Wald, see \cite{hw,bergstrom}.
In inhomogeneous models, there
will not be a one-to-one correlation between the redshift and the distance since
gravitational lensing will cause a dispersion in the Hubble diagram.
Figure~\ref{fig:sigma} shows one simulated data set of angular-diameter distances
together with the DR angular-diameter distance for three different values of 
$\alpha$.
Using $\chi^2$-tests, we determine the best-fit $\alpha$-value for 
each of our simulated data sets.
 
The case closest to the premises of the derivation of the DR distance-redshift
relation is the
case with one component of the matter density homogeneously distributed and another 
in very compact objects, e.g., point-masses. 

A perhaps more realistic model of our Universe has one part of the total
matter density in compact objects and another
part in some realistic galaxy dark matter halo 
model, e.g., the Navarro-Frenk-White
(NFW) density profile \cite{nfw}. Note that the exact parametrization of the galaxy
density profile does not significantly affect the results, see \cite{bergstrom}.

We have used three different sets of cosmological
parameter values; one open with $(\om ,\ola)=(0.2,0)$ and two flat models with    
$(\om ,\ola)=(0.3,0.7)$ and $(\om ,\ola)=(1,0)$, respectively. 

\begin{figure}[t]
  \centerline{\hbox{\epsfig{figure=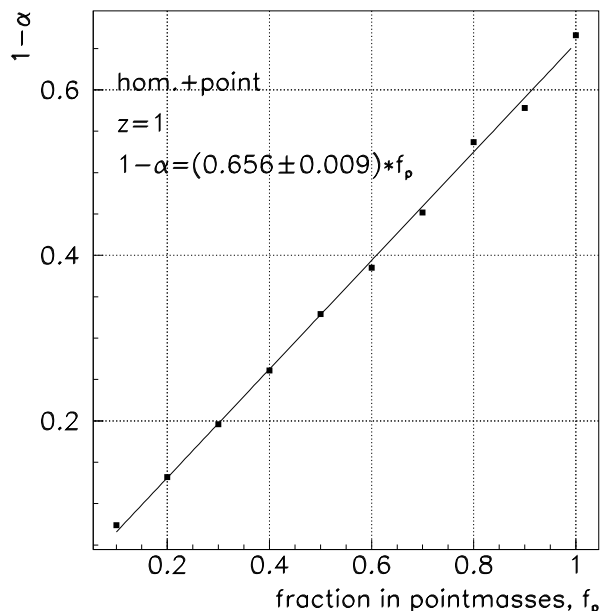,width=0.5\textwidth}}}
  \caption{Results for the homogeneity-parameter $\alpha$ 
    for the case with one homogeneous component and one 
    component in point-masses for $z=1$ and $(\om ,\ola)=(0.3,0.7)$.}
  \label{fig:z1} 
\end{figure}

\begin{figure}[t]
  \centerline{\hbox{\epsfig{figure=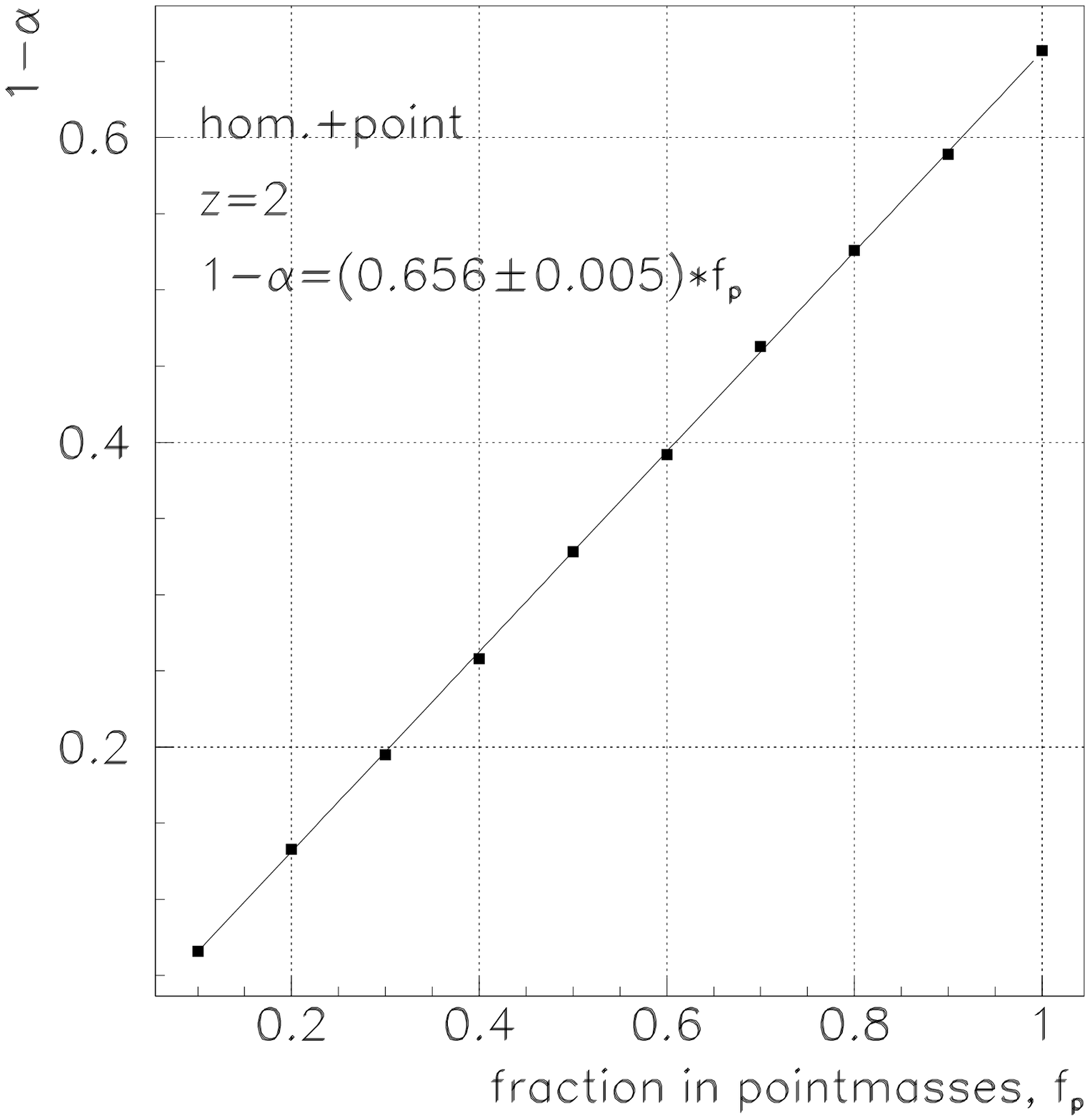,width=0.5\textwidth}}}
  \caption{Results for the homogeneity-parameter $\alpha$ 
    for the case with one homogeneous component and one 
    component in point-masses for $z=2$ and $(\om ,\ola)=(0.3,0.7)$.}
  \label{fig:z2} 
\end{figure}

\begin{figure}[t]
  \centerline{\hbox{\epsfig{figure=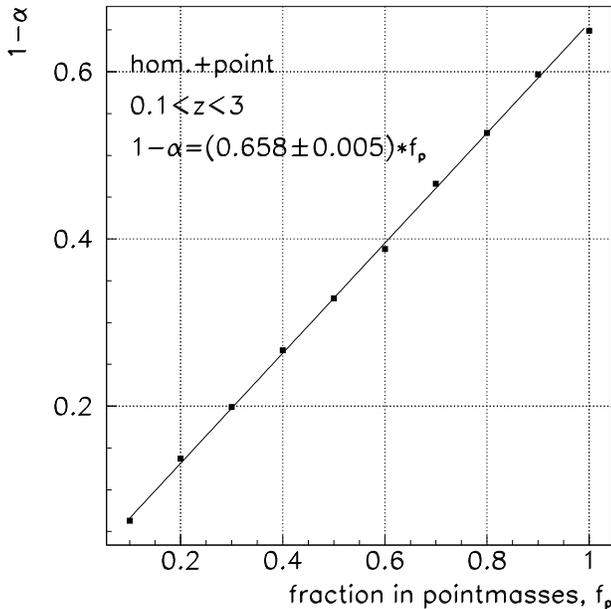,width=0.5\textwidth}}}
  \caption{Results for the homogeneity-parameter $\alpha$ 
    for the case with one homogeneous component and one 
    component in point-masses for $0.1<z<3$ and $(\om ,\ola)=(0.3,0.7)$.}
  \label{fig:zint} 
\end{figure}

In Fig.~\ref{fig:z1} to
\ref{fig:zint} we present results for $(\om ,\ola)=(0.3,0.7)$ with one homogeneous 
component and one component in point-masses. Distances are calculated 
for $z=1$, $z=2$ and a distribution of
redshifts, $0.1<z<3$. For all models and redshifts, 
there is a linear relation between the fraction of the total matter density in 
point-masses, $f_{\rm p}$, and the inhomogeneity-parameter, $1-\alpha$.
In the simulations using homogeneously distributed matter, we have added the
condition that $1-\alpha =0$ for $f_{\rm p}=0$ when fitting the linear function.

%In Fig.~\ref{fig:alpha}, we present the result of the $\chi^2$-test
%for the case of $f_{\rm p}=0.5$ and $z=1$ and $z=2$, respectively. This should be 
%compared with Fig.~15-18 and Table~IV in \cite{tomita}. 

%\begin{figure}[t]
%  \centerline{\hbox{\epsfig{figure=alpha,width=0.5\textwidth}}}
%  \caption{Results from the $\chi^2$-test with $f_{\rm p}=0.5$, 
%    $(\om ,\ola)=(0.3,0.7)$ and $z=1$ and $z=2$, respectively.}
%  \label{fig:alpha} 
%\end{figure}

\begin{figure}[t]
  \centerline{\hbox{\epsfig{figure=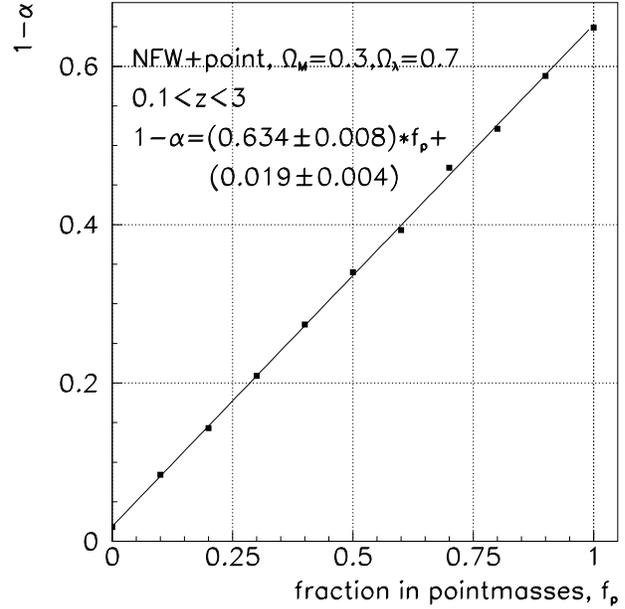,width=0.5\textwidth}}}
  \caption{Results for the homogeneity-parameter $\alpha$ 
    for the case with one component with the NFW density profile and one 
    in point-masses for $0.1<z<3$ and $(\om ,\ola)=(0.3,0.7)$.}
  \label{fig:real} 
\end{figure}

\begin{figure}[t]
  \centerline{\hbox{\epsfig{figure=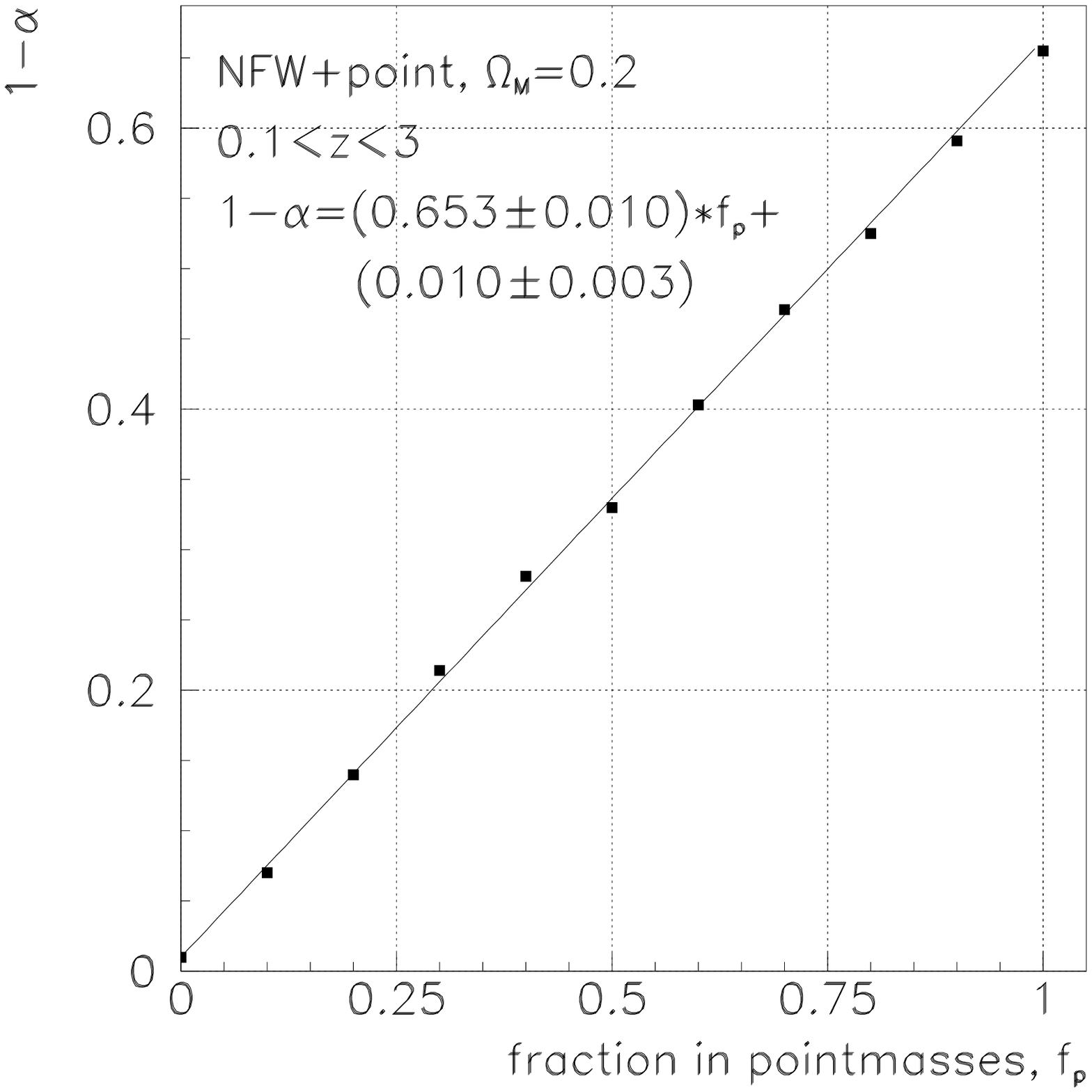,width=0.5\textwidth}}}
  \caption{Results for the homogeneity-parameter $\alpha$ 
    for the case with one component with the NFW density profile and one 
    in point-masses for $0.1<z<3$ and $(\om ,\ola)=(0.2,0)$.}
  \label{fig:o} 
\end{figure}

\begin{figure}[t]
  \centerline{\hbox{\epsfig{figure=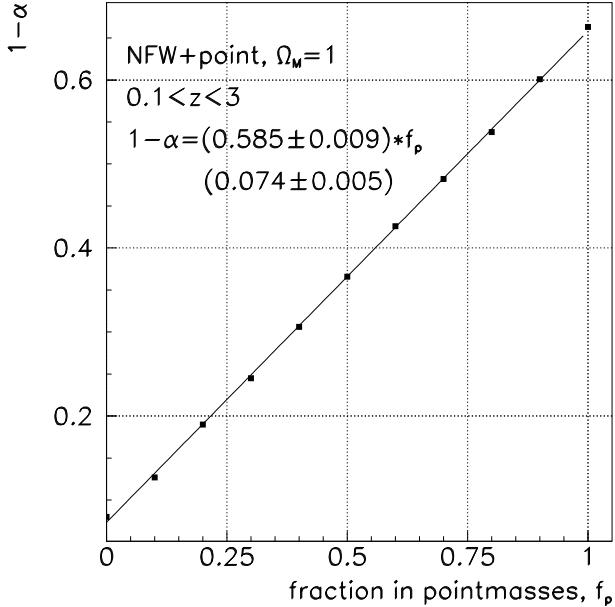,width=0.5\textwidth}}}
  \caption{Results for the homogeneity-parameter $\alpha$ 
    for the case with one component with the NFW density profile and one 
    in point-masses for $0.1<z<3$ and $(\om ,\ola)=(1,0)$.}
  \label{fig:eds} 
\end{figure}

In Fig.~\ref{fig:real} to
\ref{fig:eds}, results for the more realistic case with one part of the total
matter density in point-masses and another in dark matter halos parametrized
by the NFW density profile are presented.
Distances are calculated in a broad redshift interval $0.1<z<3$.

%==========
\section{Discussion}\label{sec:dis}
\begin{figure}[t]
  \centerline{\hbox{\epsfig{figure=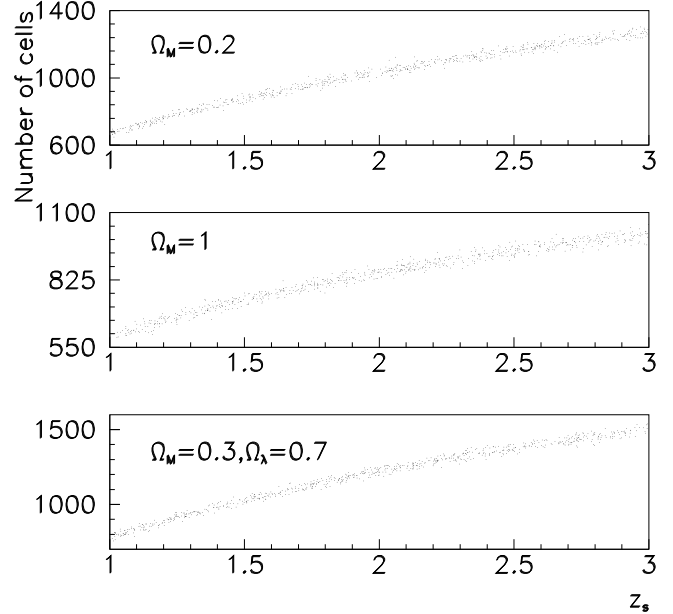,width=0.5\textwidth}}}
  \caption{Number of cells traversed in the ray-tracing simulations for 
    $(\om ,\ola)=(0.2,0)$ (top), $(\om ,\ola)=(1,0)$ (middle) and $(\om ,\ola)=(0.3,0.7)$
    (bottom).}
  \label{fig:zvscell} 
\end{figure}

For light-rays passing far from all point-masses, we expect the 
best fit value of $\alpha$ to be given by $1-\alpha\approx f_{\rm p}$ where $f_{\rm p}$ is the 
fraction of the total matter density in point-masses. Due to the magnification from 
lensing 
we expect that on average, $1-\alpha< f_{\rm p}$. It is evident from 
Fig.~\ref{fig:z1}-\ref{fig:zint} that the relation between $\alpha$ and $f_{\rm p}$ is 
independent of redshift. However, from Fig.~\ref{fig:real}-\ref{fig:eds}, we can 
see that there is a weak cosmology dependence. We can understand this effect from the
fact that lensing effects should, for a given redshift, be roughly proportional to the 
distance travelled by the light-ray as well as the mass of the lenses. 
In Fig.~\ref{fig:zvscell},
the number of cells traversed in the ray-tracing simulations are presented as a
function of redshift for the three different cosmologies used in this paper. Since the
number of cells is proportional to the distance travelled, we see that the 
$(\om ,\ola)=(0.3,0.7)$ cosmology yields the farthest distance, followed by the 
$(\om ,\ola)=(0.2,0)$ and the $(\om ,\ola)=(1,0)$ case. However, since the mass in lenses
is proportional to $\om$, we expect that we will have the largest lensing effects for 
$(\om ,\ola)=(1,0)$ followed by $(\om ,\ola)=(0.3,0.7)$ and $(\om ,\ola)=(0.2,0)$, as is
confirmed by Fig.~\ref{fig:real}-\ref{fig:eds}. 

\begin{figure}[t]
  \centerline{\hbox{\epsfig{figure=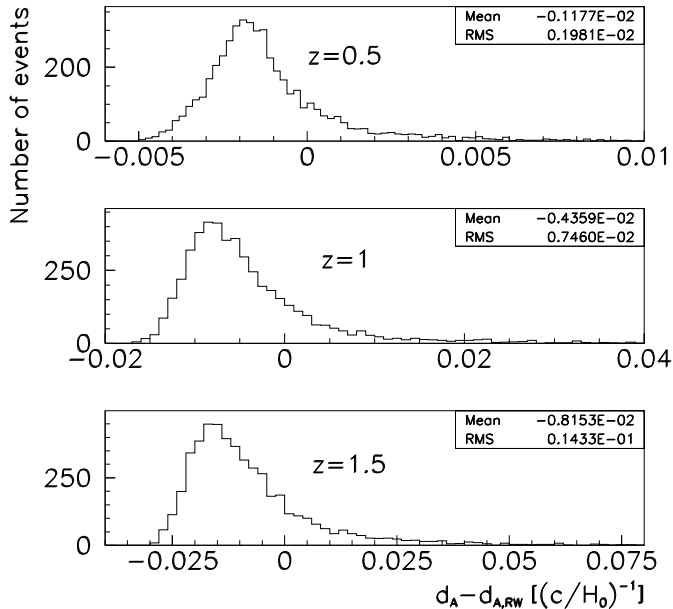,width=0.5\textwidth}}}
  \caption{The dispersion in $d_{\rm A}$ for the NFW case with 
    $f_{\rm p}=0$ and $(\om ,\ola)=(0.3,0.7)$ for $z=0.5$, $z=1$ and $z=1.5$.}
  \label{fig:rms} 
\end{figure}

We can estimate the dispersion in $\alpha$ from Fig.~\ref{fig:sigma} to be 
$\sigma_{\alpha}\approx 0.2$. However, since we are measuring distances, the 
dispersion in distance at a given redshift is perhaps of greater interest. 
In Fig.~\ref{fig:rms}, the dispersion in $d_{\rm A}$ for $f_{\rm p}=0.2$ and 
$(\om ,\ola)=(0.3,0.7)$ for $z=0.5$, $z=1$ and $z=1.5$ is shown. 
The zero-value corresponds to the value one would obtain in a homogeneous
universe, the so called Robertson-Walker angular-diameter distance, $d_{\rm A,RW}$.
The dispersion in angular-diameter distance (rms) corresponds to 
$5.9/h$ Mpc for $z=0.5$, $22/h$ Mpc for $z=1$ and
$43/h$ Mpc for $z=1.5$. It is also evident from Fig.~\ref{fig:rms} that the 
RW value is within the rms-dispersion in all three cases. However,
since the mean value is displaced from zero, the use of the RW-distance will
cause systematic errors of the order $3.5/h$ Mpc for $z=0.5$, $13/h$ Mpc for $z=1$ and
$24/h$ Mpc for $z=1.5$.  

%==========
\section{Summary}\label{sec:summary}
We have investigated the best-fit values of the homogeneity-parameter $\alpha$
in the Dyer-Roeder distance-redshift relation. For a variety of inhomogeneity models,
redshifts and cosmological parameter values, the relation between $\alpha$ and the 
fraction of compact objects, $f_{\rm p}$, is approximately linear and we can 
parametrize the relation by $$1-\alpha =a\cdot f_{\rm p},$$ where $a\approx 0.6$.
We expect the dispersion in the angular-diameter distance to be 
$5.9/h$ Mpc for $z=0.5$, $22/h$ Mpc for $z=1$ and
$43/h$ Mpc for $z=1.5$ in doing this approximation.

\section*{Acknowledgements}
The author would like to thank Ariel Goobar for useful comments.


\begin{thebibliography}{9}

\bibitem{dr}
C.C.~Dyer and R.C.~Roeder, Astrophys.~J. {\bf 180}, 31 (1973).

\bibitem{helbig}
R.~Kayser, P.~Helbig, T.~Schramm, 
Astron.~Astrophys. {\bf 318}, 680 (1995).

\bibitem{tomita}
K.~Tomita {\em pre-print} astro-ph/9806047 (1998).

\bibitem{bergstrom}
L.~Bergstr\"om, M.~Goliath, A.~Goobar, and E.~M\"ortsell, 
Astron.~Astrophys. {\bf 358}, 13 (2000).

\bibitem{hw}
D.E.~Holz, and R.M.~Wald, Phys.~Rev.~D {\bf 58}, 063501 (1998).

\bibitem{nfw}
J.F.~Navarro, C.S.~Frenk, and S.D.M.~White, Astrophys.~J. {\bf 490}, 493 (1997).

\end{thebibliography}
\end{document}